\documentclass[oldversion, useAMS]{aa} 
\usepackage{amsmath}
\usepackage{amssymb}
\usepackage{natbib}
\bibpunct{(}{)}{;}{a}{}{,}
\usepackage{graphicx}
\usepackage{txfonts}
\usepackage[english]{babel}
\usepackage{hyperref}
\usepackage{units}
\hypersetup{colorlinks=true, linkcolor=blue, citecolor=blue, urlcolor=blue}
\begin{document}
\title{Imaging the circumstellar environment of the young T Tauri star SU Aurigae}
\titlerunning{Imaging SU Aur at the T Tauri/Herbig transition}


\author{
	S.~V.~Jeffers\inst{1,2}
	\and
	M.~Min\inst{3,2}
	\and
	H.~Canovas\inst{4}
	\and
	M.~Rodenhuis\inst{2}
	\and
	C.~U.~Keller\inst{2} 
	\thanks{Based on observations made with the William Herschel Telescope operated on the island of La Palma by the Isaac Newton Group in the Spanish Observatorio del Roque de los Muchachos of the Instituto de Astrofísica de Canarias}}

\offprints{S.~V. Jeffers, \email{Jeffers@astro.physik.uni-goettingen.de}}

\institute{Institut f\"{u}r Astrophysik, Georg-August-Universit\"{a}t, Friedrich-Hund-Platz 1, 37077 G\"{o}ttingen, Germany
\and
Leiden Observatory, P.O. Box 9513, NL-2300 RA  Leiden, The Netherlands 
\and
Sterrenkundig Instituut Anton Pannekoek, Universiteit van Amsterdam, Postbus 94249, 1090 GE Amsterdam, The Netherlands 
\and
Departamento de Fisica y Astronomia, Universidad de Valparaiso, Valpariso, Chile}

   \date{\today}

  \abstract
   {The circumstellar environments of classical T Tauri stars are challenging to directly image because of their high star-to-disk contrast ratio. One method to overcome this is by using imaging polarimetry where scattered and consequently polarised starlight from the star's circumstellar disk can be separated from the unpolarised light of the central star.  We present images of the circumstellar environment of SU Aur, a classical T Tauri star at the transition of T Tauri to Herbig stars.  The images directly show that the disk extends out to ~500 au with an inclination angle of $\sim$ 50$^\circ$.  Using interpretive models, we derived very small grains in the surface layers of its disk, with a very steep size- and surface-density distribution.  Additionally, we resolved a large and extended nebulosity in our images that is most likely a remnant of the prenatal molecular cloud.  The position angle of the disk, determined directly from our images, rules out a polar outflow or jet as the cause of this large-scale nebulosity.}

\keywords{circumstellar matter -- dust, extinction, stars -- individual (SU Aurigae), stars -- mass loss}

   \maketitle

\section{Introduction}

Classical T Tauri stars represent the earliest evolutionary stages of
low- to intermediate-mass star formation that are optically visible.  They are typically characterised by a rotating disk of gas and dust and are the low-mass counterparts to the Herbig stars typically having spectral types from late F to M.  During a star's T Tauri phase its circumstellar disk is the source of material that is accreted onto the star and provides the building blocks from which planets can form \citep[e.g.][and references therein]{williams2011}. To understand how planets could form in the disks of T Tauri stars, it is first necessary to understand the large-scale shape, structure, and composition of their disks. \\

The direct imaging of circumstellar environments is challenging
because they are often fainter by many orders of magnitude fainter than their central stars. A powerful technique that overcomes this is polarimetric imaging. This technique is based on the physics that light reflected
by circumstellar material becomes linearly polarised in the reflection
process and can be easily separated from the unpolarised light of the
central star.  Polarimetric imaging has been successfully applied to the imaging of
 the circumstellar environments of many young stars, with recent
highlights including the Herbig stars HD 169142 \citep{quanz2013}, HD 142527, \citep{Canovas2013} and AB Aur \citep{hashimoto2011} where annular gaps and inner holes are likely to have been carved out by orbiting protoplanets.  Additionally, polarimetry is
advantageous because of its ability to determine much-needed
additional constraints such as the surface-grain composition, size, and radial distribution. Knowing these parameters greatly improves
the reliability of the model-fitting procedure than could ever be
determined from intensity images alone or from the star's spectral energy distribution (SED).\\

Currently, disks have been resolved for $\approx$ 50 T Tauri stars predominately around stars with late-K and M spectral types \footnote{www.circumstellardisks.org}.  We present the first direct images of the disk around the G2 classical T Tauri star SU Aurigae, a circumstellar disk at the T Tauri/Herbig Ae transition, using imaging polarimetry.  The observations were secured using ExPo, the EXtreme POlarimeter \citep{rodenhuis2012}, which is a regular visitor-instrument at the 4.2m William Herschel Telescope.  The layout of the paper is as follows: in Section 2 we summarise the current knowledge on SU Aur, in Section 3 we describe the observational procedure and the data analysis, with the polarimetric image of SU Aur presented in Section 4. The observations are interpreted using radiative-transfer modelling in Section 5, and the results are discussed in Section 6.  We present our conclusions in Section 7.  

\section{Target: SU Aurigae}

SU Aurigae is a classical T Tauri star \citep{Herbig1952,giampapa1993,bouvier1993} and is a member of the Taurus-Auriga star-forming region, with a distance of 143 $^{+17}_{-13}$ pc \cite{Bertout2006}.  It is  classified as G2 subgiant with an exceptionally fast rotation v sin i = 66 km s$^{-1}$ and is one of the brightest classical T Tauri stars with V=+9.16 mag and (B-V)=+0.90.  SU Aur is also characterised by strong photometric variability caused by dark starspots and bright accretion hotspots \citep{dewarf2003}.


SU Aur has been the target of many studies at all wavelength ranges,  though these were primarily focused on understanding the central star and not its large-scale circumstellar environment.  Observational signatures of a circumstellar disk are present in SU Aur's infrared and UV excess \citep{Bertout1988,kenyon1995} and in spectroscopic observations that show optical veiling from circumstellar and/or accreting material \citep{unruh2004}.

However, the most detailed observations of SU Aur's circumstellar disk have been secured using long-baseline interferometry from the Palomar Testbed Interferometer \citep{akeson2002,akeson2005}.  These observations are most sensitive to the hot-disk material that is close to the star, resulting in a reliable determination of the inner disk radius (0.13/0.18 au depending on whether the model is face-on or inclined \citep{akeson2002}).  This value agrees with the spectropolarimetric observations of H$\alpha$ reported by \citep{vink2005} who also suggest that SU Aur is surrounded by an optically thick scattering disk.   Additionally, \cite{akeson2002,akeson2005} determined SU Aur's inclination to be 62$^\circ$ which agrees with the value of $\approx$60$^\circ$ derived by \cite{unruh2004} from tomographic Doppler imaging of starspots on the central star.  A summary of SU Aur's stellar parameters is given in Table~\ref{tab:st_param}.



\begin{table}
\caption{SU Aur's stellar parameters}
\protect\label{tab:st_param}
\begin{tabular}{l c c}
\hline
\hline
Parameter &  & Reference \\
\hline
Coordinates$^a$
RA & 04 55 59.39 &  \\
\hspace{1.5cm} DEC & +30 34 01.52 & \\
Spectral type & G2 III & 1 \\ 
M$_V$ & 9.16 mag &  2 \\
B-V 	& +0.90 & 2 \\
Effective temperature & 5860$\pm$100K & 3 \\
Mass & 1.88$\pm$0.1 M$_\odot$ & 3 \\
Radius & 3.5 R$_\odot$ & 4 \\
v sini & 66 km s$^{-1}$ & 5 \\
       &                & 6 \\
Age (log t) & 6.8$\pm$0.08 yr & 7 \\
Luminosity & 9.29$^{2.26}_{-1.65}$  L$_\odot$ & 3 \\
 & $\sim$9.7 L$_\odot$ &  2$^b$ \\
Inclination & 62 $^{+4}_{-8}$ degrees & 8 \\
Distance & 143 pc & 9 \\
A$_v$ & 0.9 mag & 3 \\
\hline 
\hline
\end{tabular}

 (1) \cite{Herbig1952};(2) \cite{dewarf2003}; (3) \cite{Bertout2007}; (4) \cite{muzerolle2003}; (5) \cite{johnskrull1996}; (6) \cite{unruh2004}; (7) \cite{Bertout2007}; (8) \cite{akeson2002}; (9) \cite{Bertout2006}.
$^a$ The stellar coordinates are taken from Simbad (http://simbad.u-strasbg.fr). 
$^b$ The value of the luminosity of \cite{dewarf2003} comprises the luminosity of the  star = 6.3$\pm$1.6 L$_\odot$ and the luminosity of the disk $\sim$3.4 L$_\odot$. 
\end{table}

\begin{figure*} 
\begin{center}
\includegraphics[scale=0.65]{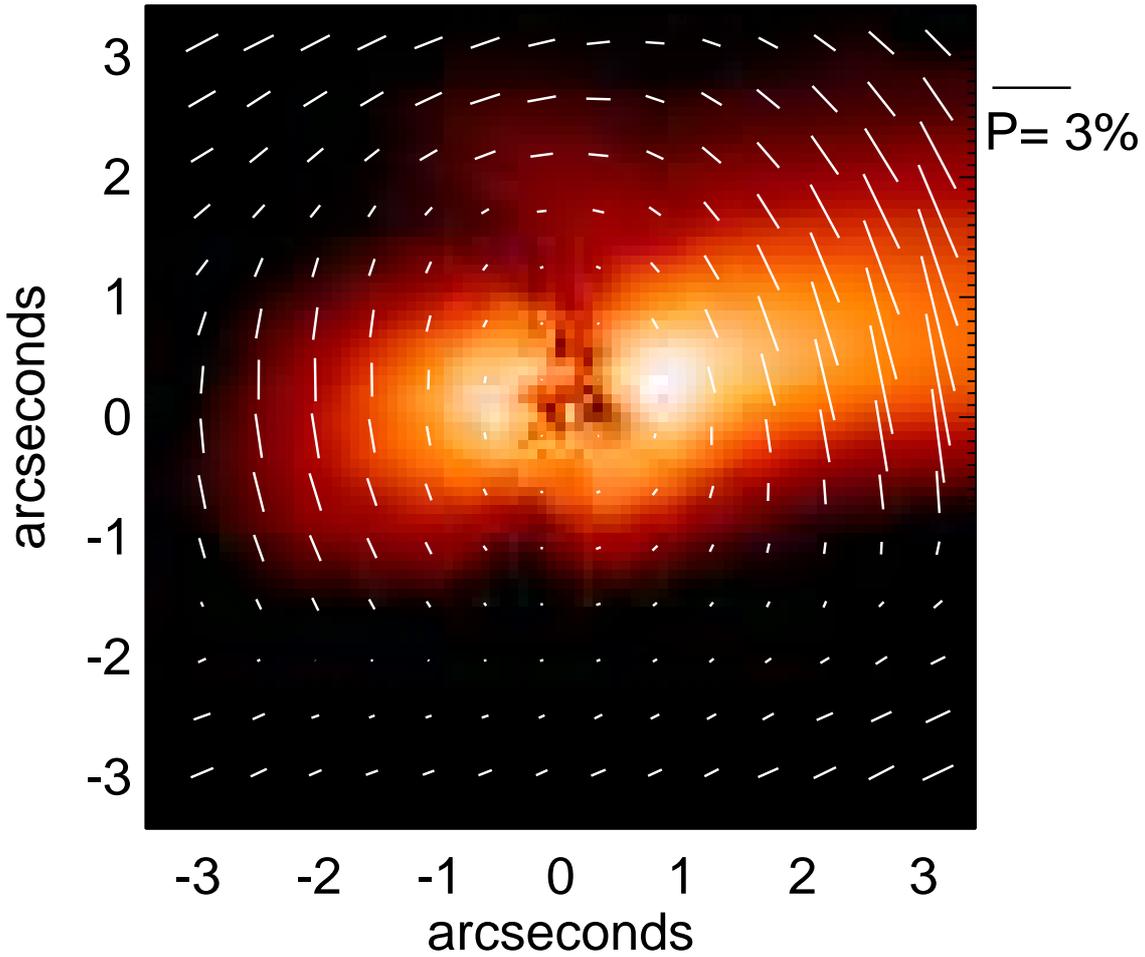}
\caption{ExPo image of the circumstellar environment of SU Aur in polarised intensity (P$_I
$), taken without a filter.  The intensity scale is logarithmic over two orders of magnitude and the length of the vectors represents a lower limit to the true polarisation degree.  North is up and east is to the left.} 
\protect\label{f-expoim1}
\end{center} 
\end{figure*}

\section{Observations and data analysis}

In this section, we describe the technical setup and summarise the data reduction pipeline.

\subsection{ExPo}

ExPo is a sensitive imaging-polarimeter that was designed and built by our team.  It is a regular visitor instrument at the Nasmyth focus of the 4.2~m William Herschel Telescope.  ExPo has a 20\arcsec $\times$20\arcsec field-of-view and a wavelength range of 500 to 900 nm. ExPo achieves a sensitivity in linear polarisation of 10$^{-4}$, allowing contrasts of up to 10$^{-6}$ with respect to the star at a separation of a few arcseconds \citep{rodenhuis2012}. ExPo is currently not used for observations in circular polarisation.

The polarimetric sensitivity is achieved by employing the beam-exchange method \citep{Semel1993}. This method reduces systematic effects by simultaneously detecting orthogonal linear polarisation states and exchanging the polarisation states between the two beams.  ExPo uses a polarising cube beamsplitter and a ferro-electric liquid crystal (FLC) switchable half-waveplate synchronised with the 35 Hz frame rate of the EM-CCD camera. ExPo observations comprise long sequences of modulated short exposures for each FLC position, each lasting 28.8 ms. The EM-CCD technology enables the observation of comparatively faint targets (down to 13$^{th}$ magnitude in the V-band) at these short exposure times without being dominated by detector read out noise.

For a given position of the FLC polarisation modulator, the instrument will be sensitive to polarisation in two orthogonal directions (e.g. Stokes Q), but not in the 45$^\circ$ direction (Stokes U). To measure polarisation along those axes, the FLC is rotated by 22.5$^\circ$. For redundancy, data are taken at four FLC angles (0$^\circ$, 22.5$^\circ$, 45$^\circ$ and 67.5$^\circ$).

The observations of SU Aur were obtained without the use of an adaptive optics system. Our data analysis method (described below) makes use of the numerous short exposures by including a custom shift-and-add image combination approach \citep{Canovas2011}, which reduces the effect of atmospheric seeing.

\subsection{Observations}

The observations of SU Aur were secured as part of a larger observing campaign to image the circumstellar environments of young stars from 27 December 2010 to 4 January 2010.  To measure the complete linear polarisation signal (Stokes Q and U) four subexposures of 4000 frames each were taken per FLC position. The observations of SU Aur are summarised in Table ~\ref{tab:obs}.  The polarised features described in this paper are present in both data sets, though only the data set from 1 January is used in this analysis as the atmospheric seeing was much lower.

For ExPo, the FLC modulates the orientation of the linear polarisation state of the incoming light by 90$^\circ$ every 0.028s switching between successive A and B frames, resulting in a total of 2000 A + 2000 B frames = 4000 frames.  For each frame (A or B) the light is separated using a beamsplitter into two orthogonal beams (left and right beams), which are simultaneously imaged on the EM-CCD, a concept that is the basis of a dual-beam polarimeter.  For every pair (A and B frames) of observations, this results in four different images of SU Aur (with a total of 8000 for each FLC angle as described above), for instance (A+B): $A_\mathrm{left}, A_\mathrm{right}, B_\mathrm{left}, B_\mathrm{right}$.

The total exposure time of each subexposure is 4000 * 0.028~s (total=112~s). At the end of every FLC sequence a series of dark frames are taken per target,  comprising approximately 8000 frames.  At the beginning and end of every night a set of dome flat-fields were taken with a polarisation calibrator. Approximately 1000 dome flats were combined to produce one masterflat per FLC position (0$^\circ$, 22.5$^\circ$, 45$^\circ$ and 67.5$^\circ$).

\subsection{Data reduction}

The data-reduction software has specifically been developed for ExPo with the aim of minimising atmospheric and instrumental effects.  The aim is to separate the linearly polarised signal from that of the unpolarised starlight.  The first step is the processing of individual frames, then the four subframes ($A_\mathrm{left}, A_\mathrm{right}, B_\mathrm{left}, B_\mathrm{right}$) are combined in sequence to form a polarised intensity image.  This image is then combined at four FLC angles with calibration data to form a calibrated polarised intensity image.  The data analysis is described in more detail by \cite{Canovas2011} and is briefly summarised here. 

\subsubsection{Individual frame processing}

Each image is first individually corrected for dark, flatfield and bias effects.  Cosmic rays are identified and removed by comparing the two simultaneous left and right images of each frame.  Finally, the frames are corrected for streaking caused by frame transfer or read-out of the detector.

\subsubsection{Instrumental polarisation}

Because ExPo is mounted on a Nasmyth platform, the third mirror of the telescope can introduce a considerable amount of polarisation cross-talk, in particular $I \rightarrow Q, U$.  Consequently, ExPo has been designed 
to limit the sources of instrumental polarisation and to remove instrumental polarisation by the use of a polarisation compensator, as described by \cite{rodenhuis2012}.  Any remaining instrumental polarisation is removed during the data-reduction process.  Both steps rely on the assumption that the central star is unpolarised, an assumption that has also been extensively used \citep[e.g.][]{perrin2008,quanz2011,Canovas2012,Jeffers2012}. The resulting values of polarisation should be treated as lower limits of the true polarisation signal of the disk and circumstellar environment of SU Aur, as discussed by \cite{min2012}.

\subsubsection{Combining images}

The first step is to align the images by applying a two-step image-alignment algorithm. First a (subset of) the images are aligned with respect to their brightest pixel to create a template image. The images are then aligned once more with subpixel accuracy through cross-correlation with this template. \cite{Canovas2011} showed that this procedure is more effective for ExPo data than the classical shift-and-add method.

Because ExPo is a dual-beam polarimeter, \cite{Canovas2011} showed that for ExPo the best method for combining the two beams is by image differences.  Other examples of combining the two beams can be found in \cite{kuhn2001}, \cite{perrin2008}, \cite{quanz2011}, and \cite{Hinkley2009}.  A linearly polarised image is obtained from the double-difference equated as

\begin{equation}
P'_l=0.5(\Delta A-\Delta B)=0.5((A_\mathrm{left}-A_\mathrm{right})-(B_\mathrm{left}-B_\mathrm{right})).
\end{equation}

The total intensity image (disk + central star) is derived from

\begin{equation}
I=0.5(A_\mathrm{left}+A_\mathrm{right}+B_\mathrm{left}+B_\mathrm{right}).
\label{eq_dd}
\end{equation}

ExPo does not use a derotator. Therefore, the combination is not performed for the complete sequence at once, but in blocks that are short enough to limit rotation effects within the block. Each block is then derotated before the final combination. This is part of the polarisation calibration.

\subsubsection{Polarisation calibration}

The final step of the data reduction process combines a minimum of two sequences, taken at FLC angles 22.5$^\circ$ apart, with calibration measurements to produce a calibrated image in three Stokes parameters: I, Q, and U. Based on the double-difference polarisation signals, we can define a modified signal matrix $\mathsf{X}$ as follows:

\begin{equation}
\binom{{P^{\prime}}_\mathrm{FLC1}}{{P^{\prime}}_\mathrm{FLC2}} = X \cdot \binom{Q}{U},
\end{equation}

where ${P^{\prime}}_\mathrm{FLC1}$ is the result of Equation~\ref{eq_dd} for the first FLC angle and ${P^{\prime}}_\mathrm{FLC2}$ for the second FLC angle, $X$ is the singular matrix that relates the measured signal to the input Stokes Q and U vectors.  The parameters of the matrix $\mathsf{X}$ are obtained by taking dome-flat measurements using a linear calibration polariser. A calibration set consisting of 20 dome-flat sequences (4 FLC angles + 1 without FLC $\times$ 4 calibration polariser positions) is obtained either at the start or at the end of every observing night. The parameters of $\mathsf{X}$ are obtained by fitting the response parameters $a$ and $b$ for each FLC angle:

\begin{equation}
P^{\prime}_{\mid} = (a\,\,b) \times \binom {Q}{U},
\end{equation}

The positions of the calibration polariser define the polarisation reference frame of the Stokes Q and U parameters. However, the calibration polariser will not produce 100\% Stokes Q or U polarised light: The dome-flat lamps may be polarised and the Nasmyth tertiary certainly introduces polarisation. A bootstrap method is therefore used where the telescope beam is initially assumed to be unpolarised and a first estimate of the signal matrix is obtained as outlined above. This is used to calibrate a dome flat taken without the calibration polariser to estimate the actual polarisation of the telescope beam. The Stokes Q \& U values in the equations above are then corrected and the process is repeated until the estimate of the input polarisation no longer changes. This occurs in two to three iterations and produces a calibration accuracy at the 1\% level. The signal matrix $\mathsf{X}$ obtained can now be inverted to produce a calibration matrix that is used to generate the images in Stokes Q and U from the double-difference polarisation data. The procedure is described in more detail by \cite{rodenhuis2012}.

The polarisation images of SU Aur presented here show the polarised intensity, defined as $P_{I}=\sqrt{Q^2+U^2}$, overlaid with vectors representing the polarisation orientation, obtained from ${\Theta}_P=0.5 \times \arctan (U/Q)$. The vectors are scaled to represent the degree of polarisation, defined as $P_{I}/I$, where I includes the contribution from both the star and the disk. Therefore, the length of the vectors represents a lower limit of the true polarisation degree of the disk. 

\begin{table}
\caption{Summary of ExPo observations of SU Aur.}
\protect\label{tab:obs}
\begin{tabular}{l c c c c}
\hline
\hline
Instrument & Date & Wavelength & Exp time(s) & No exp$^1$ \\
\hline
ExPo & 27 December 2009 & 500-900nm & 0.028 & 4x4000\\ 
 & 1 January 2010 & 500-900nm & 0.028 & 4x4000 \\ 
\hline
\hline
\end{tabular}
$^1$ Each exposures comprises images taken at four FLC angles: 0$^\circ$, 22.5$^\circ$, 45$^\circ$ and 67.5$^\circ$.
\end{table}

\subsection{Errors}

The polarisation degree derived from the ExPo images should be taken as a lower limit, since we cannot measure the total intensity from the disk alone. The main uncertainties in the polarised intensity images are from the calibration of the images, and from the instrumental polarisation correction, which are lower than 2\% \citep{rodenhuis2012} to within $\sim$0.8".  At larger distances from the star $\gtrsim$ 3" and for stars with a significantly different spectral type to the calibration lamps, these values will be much higher.   These uncertainties do not affect the shape and extent of the disk, but only the absolute values of P, which are not used as constraints on the interpretive disk models. 

\section{Polarimetric image of SU Aur}

The resulting ExPo image of SU Aur in linearly polarised light is shown in Figure~\ref{f-expoim1}.  The image is zoomed-in to highlight the structure observed in the inner parts of the ExPo image.  The radial extent of the disk is at least $\sim$2", which at SU Aur's distance translates into a size of a few hundred au.  The inclination of the disk can be constrained to first order via the ratio between the width (the major axis of the projected ellipse) of the disk (2") and the height (the minor axis of the projected ellipse) of the disk (1").  This implies that the disk's inclination is higher than 30$^\circ$ from the pole.  From the polarimetric models of \cite{min2012}, it is clear that SU Aur's disk is a flaring disk because the very bright regions of the two bright white east and west lobes are only a small fraction of the total area of the resolved disk, whereas for a settled disk they are approximately half of the resolved image \citep[][top middle panel of Figure 7  for flared disks and Figure 8 for settled disks]{min2012}.  

\section{Modelling SU Aur's circumstellar environment}


\begin{figure*}
\def\imagetop#1{\vtop{\null\hbox{#1}}}
\begin{tabular}[h]{c c c}        
  \emph{Cartoon Models} & \emph{Polarised Intensity Models} & \emph{Convolved with simulated ExPo PSF} \\        
  \imagetop{\includegraphics[width=0.265\textwidth]{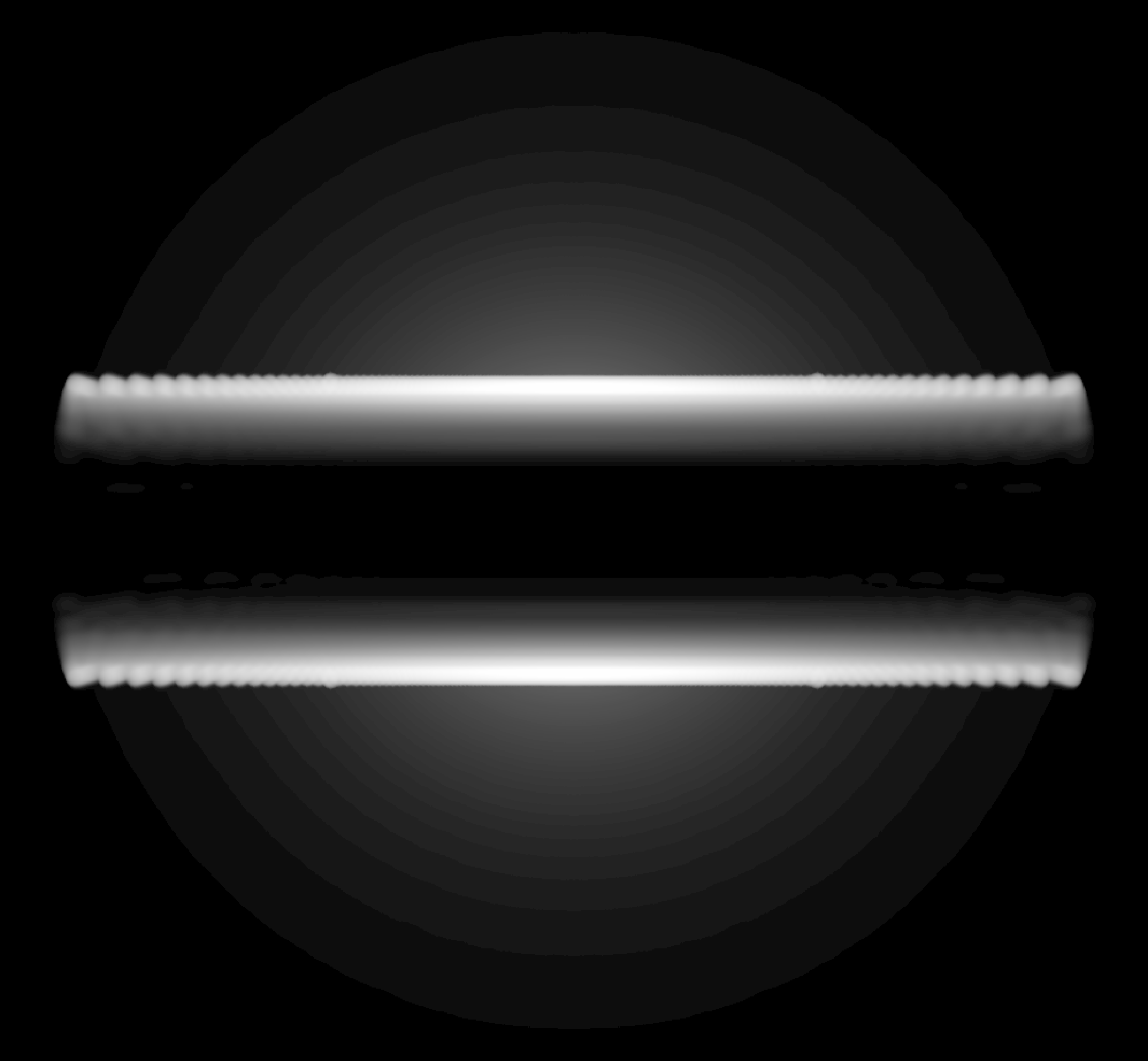}} &
  \imagetop{\includegraphics[width=0.34\textwidth]{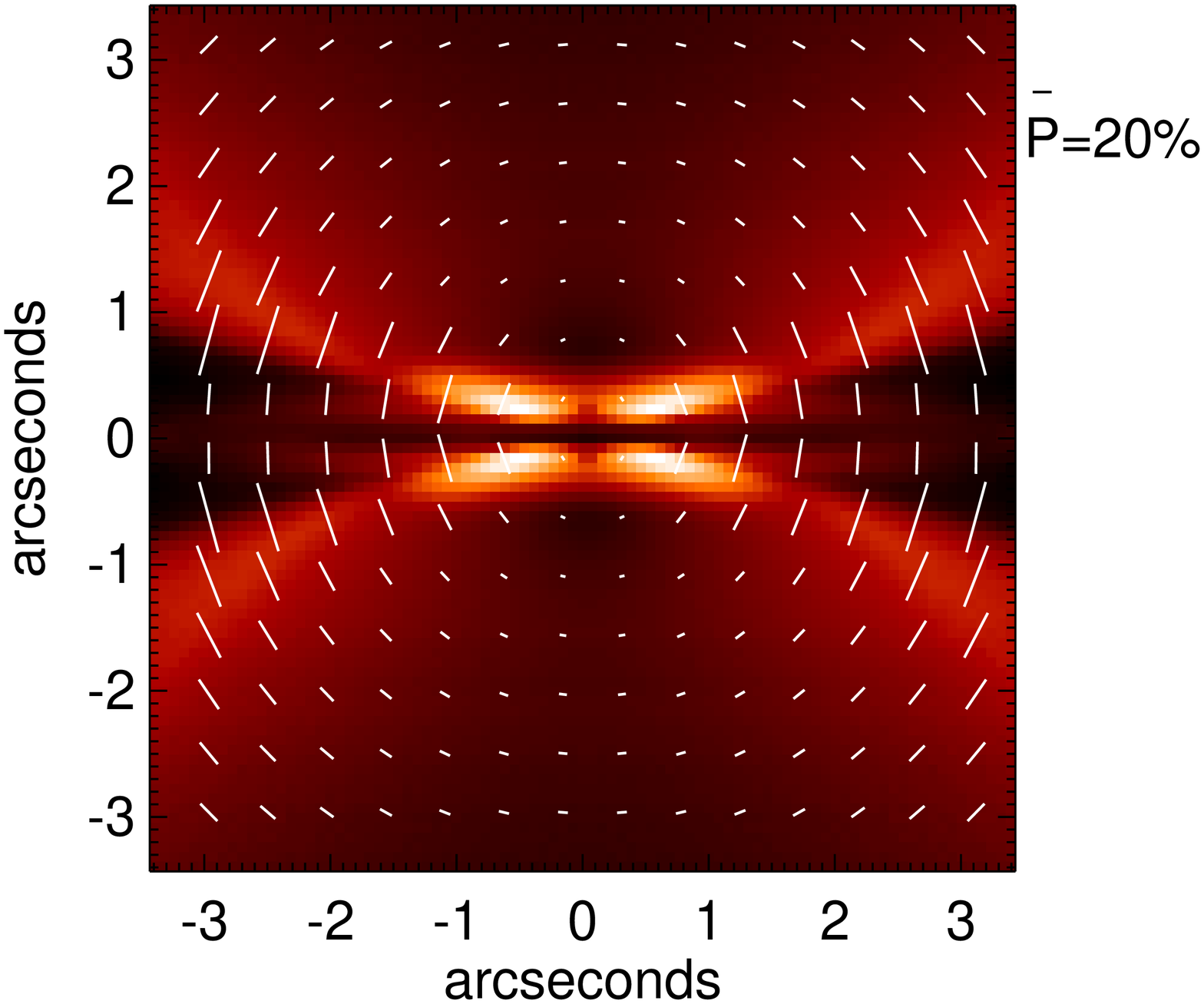}}&
  \imagetop{\includegraphics[width=0.34\textwidth]{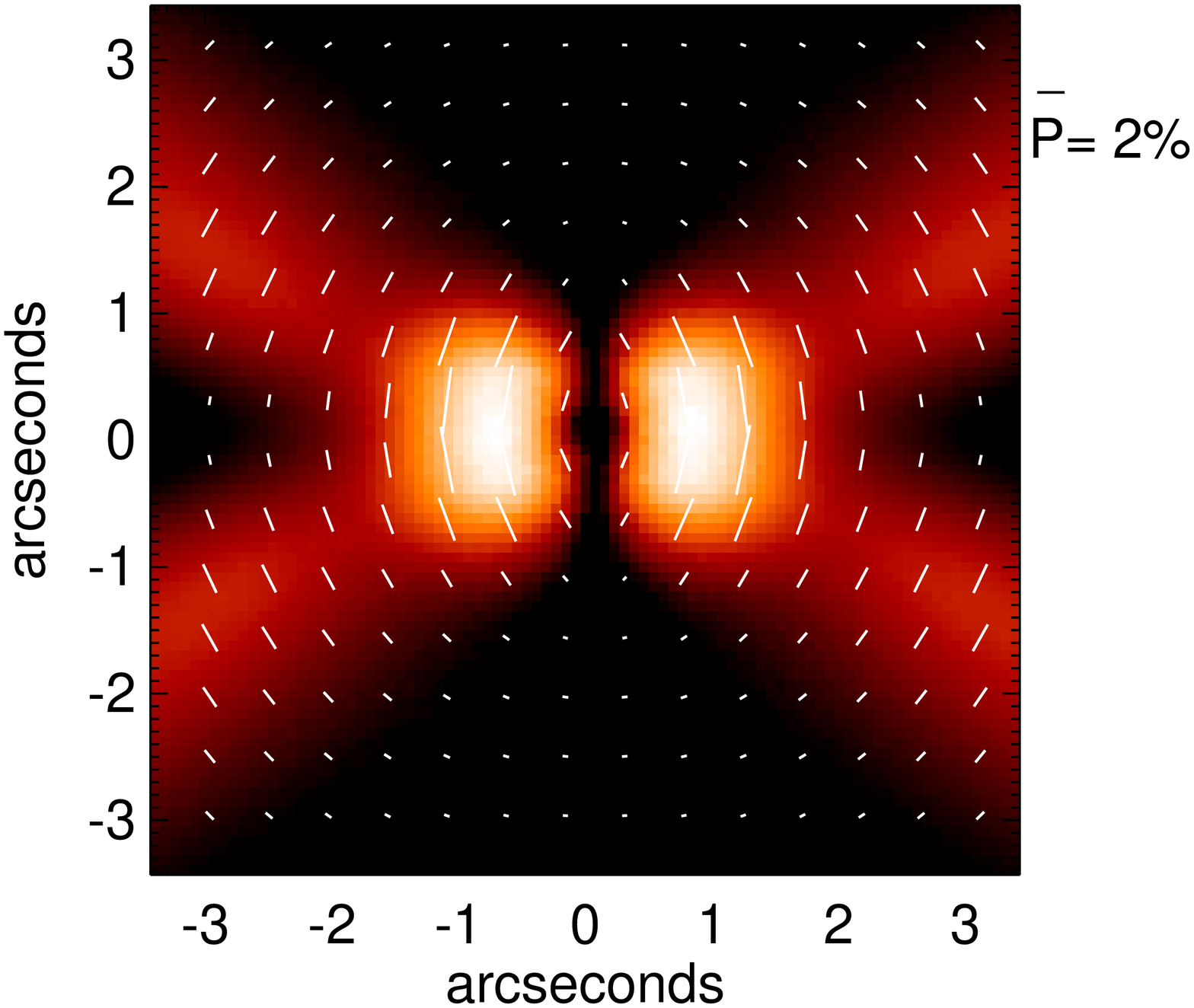}} \\
  \imagetop{\includegraphics[width=0.265\textwidth]{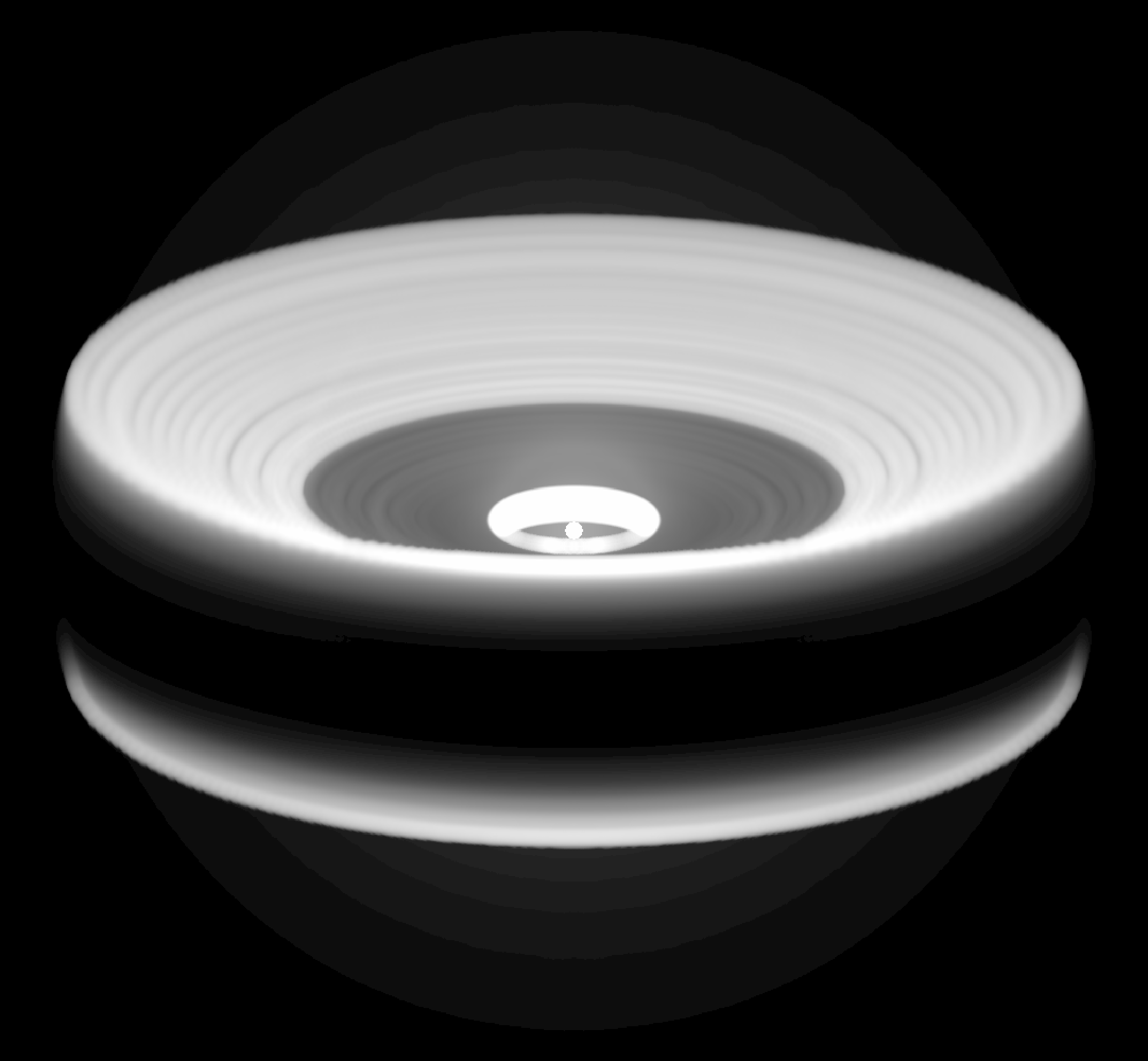}} &
  \imagetop{\includegraphics[width=0.34\textwidth]{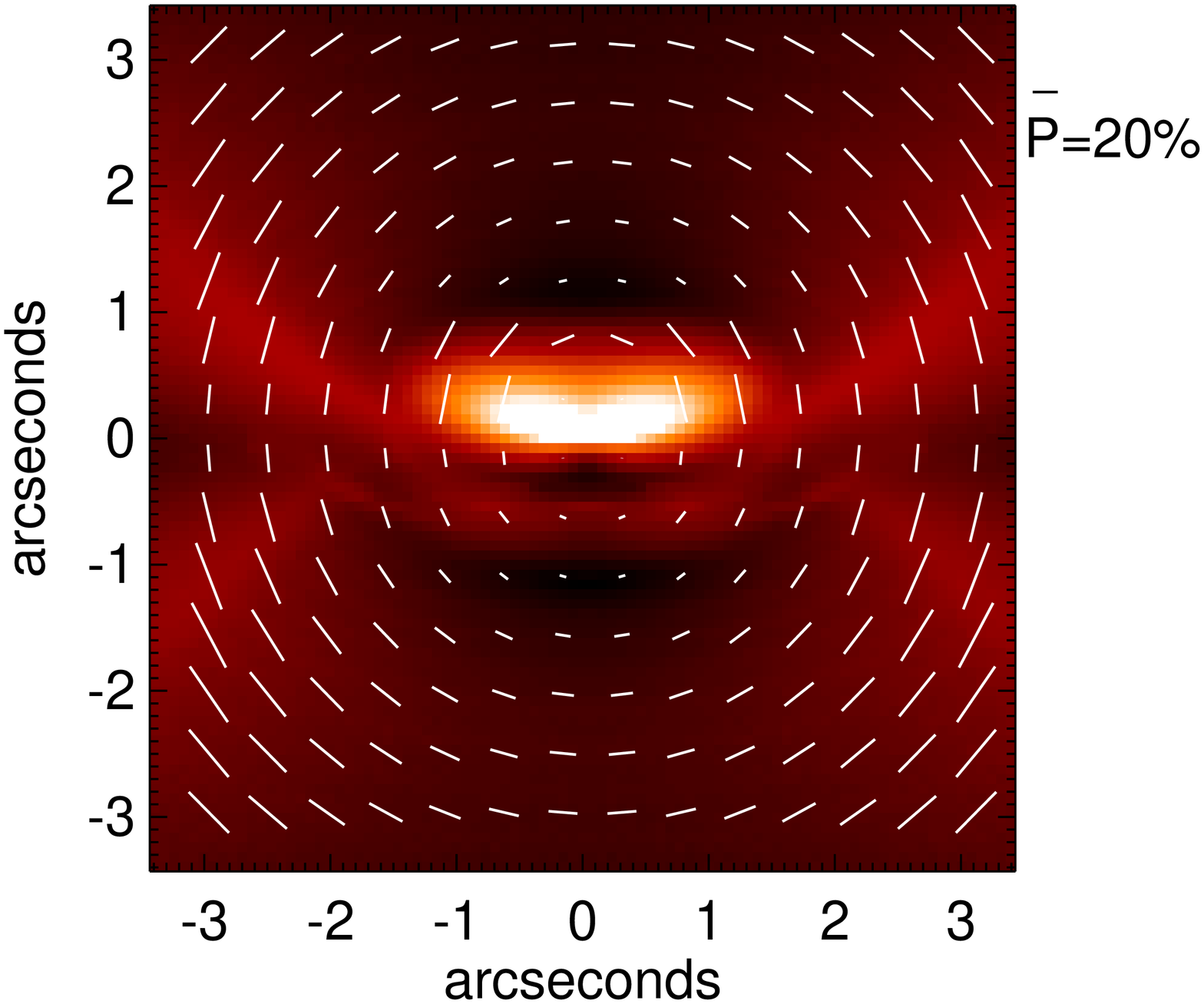}}&
  \imagetop{\includegraphics[width=0.34\textwidth]{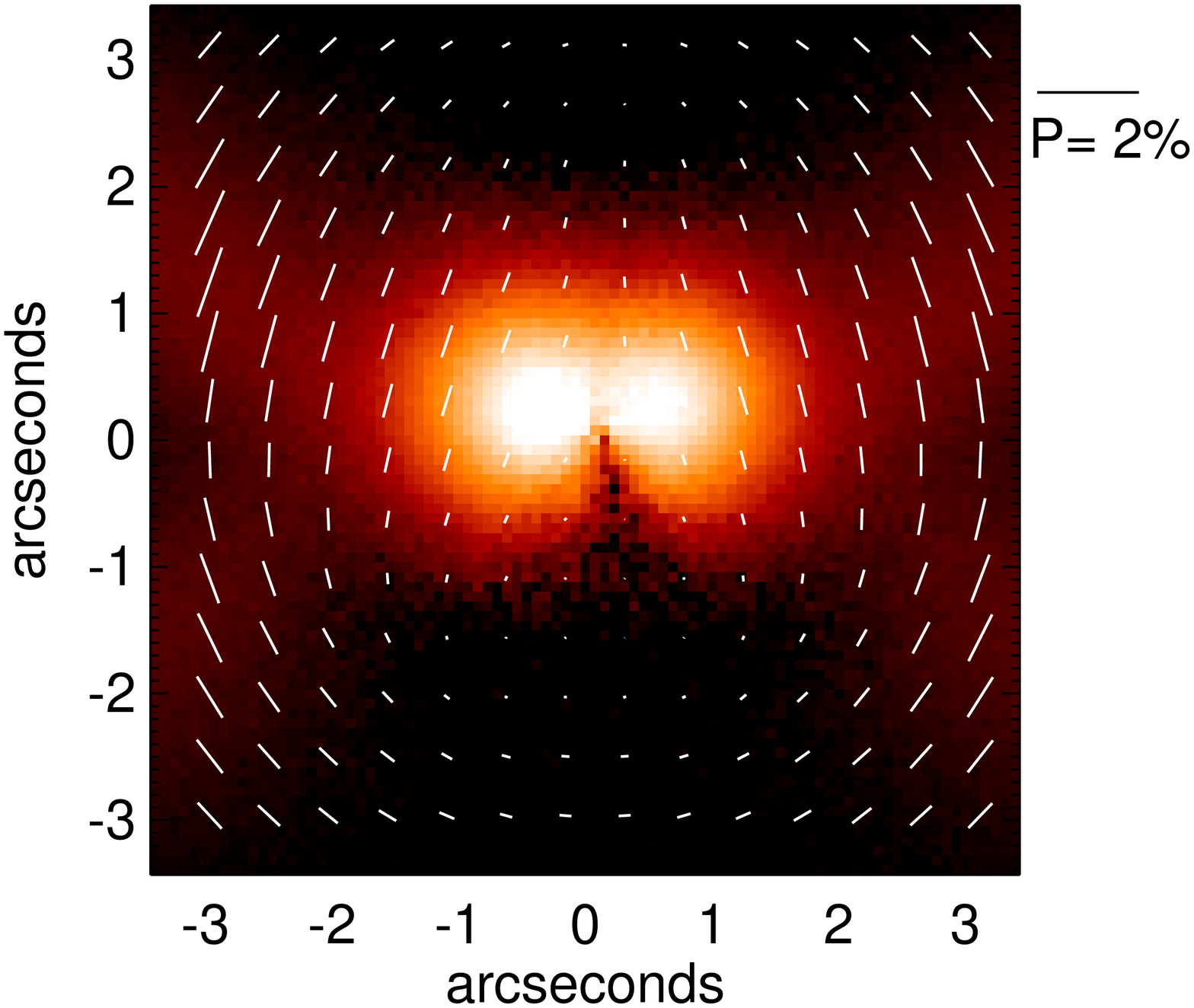}} \\
  \imagetop{\includegraphics[width=0.265\textwidth]{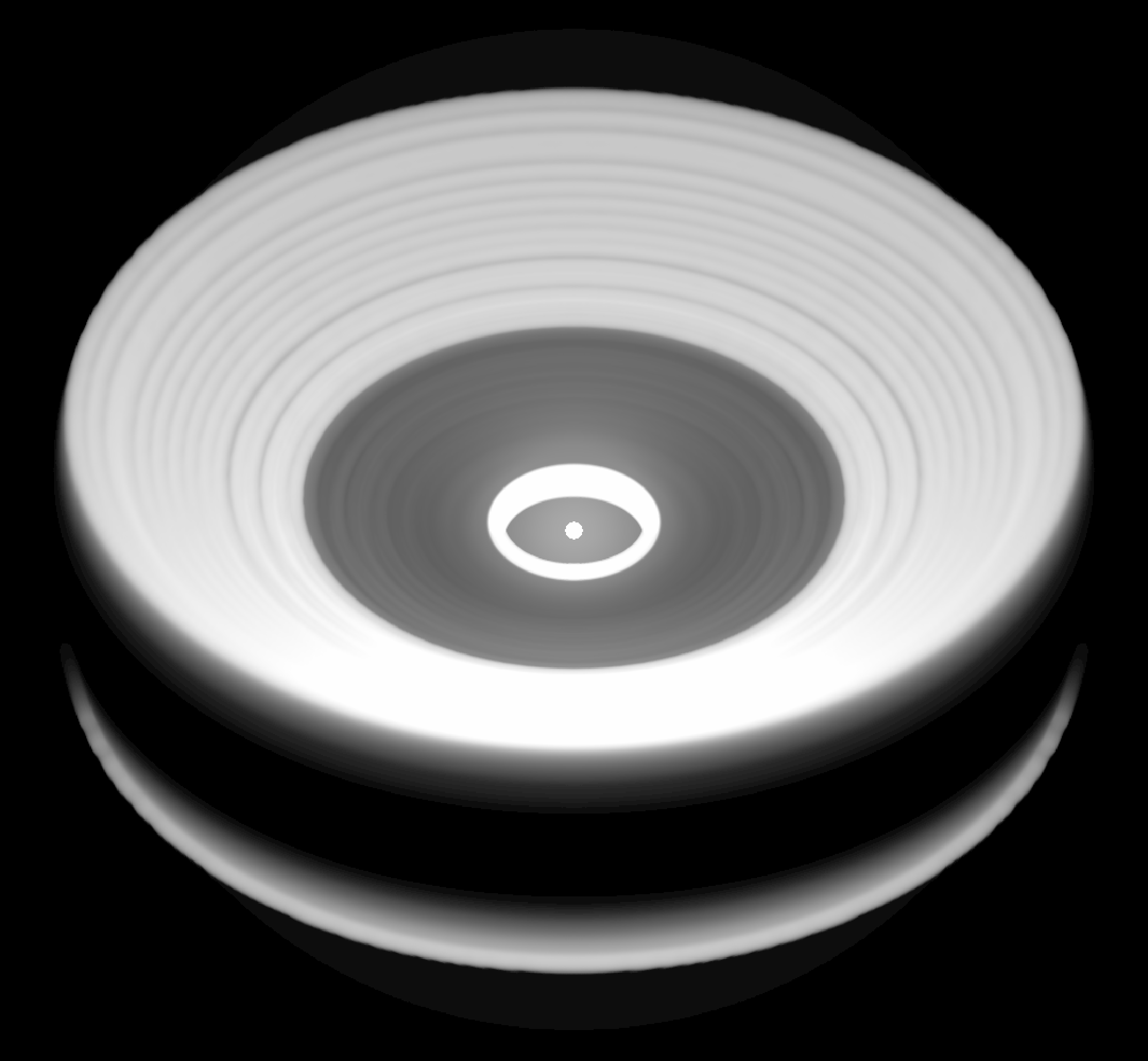}} &
  \imagetop{\includegraphics[width=0.34\textwidth]{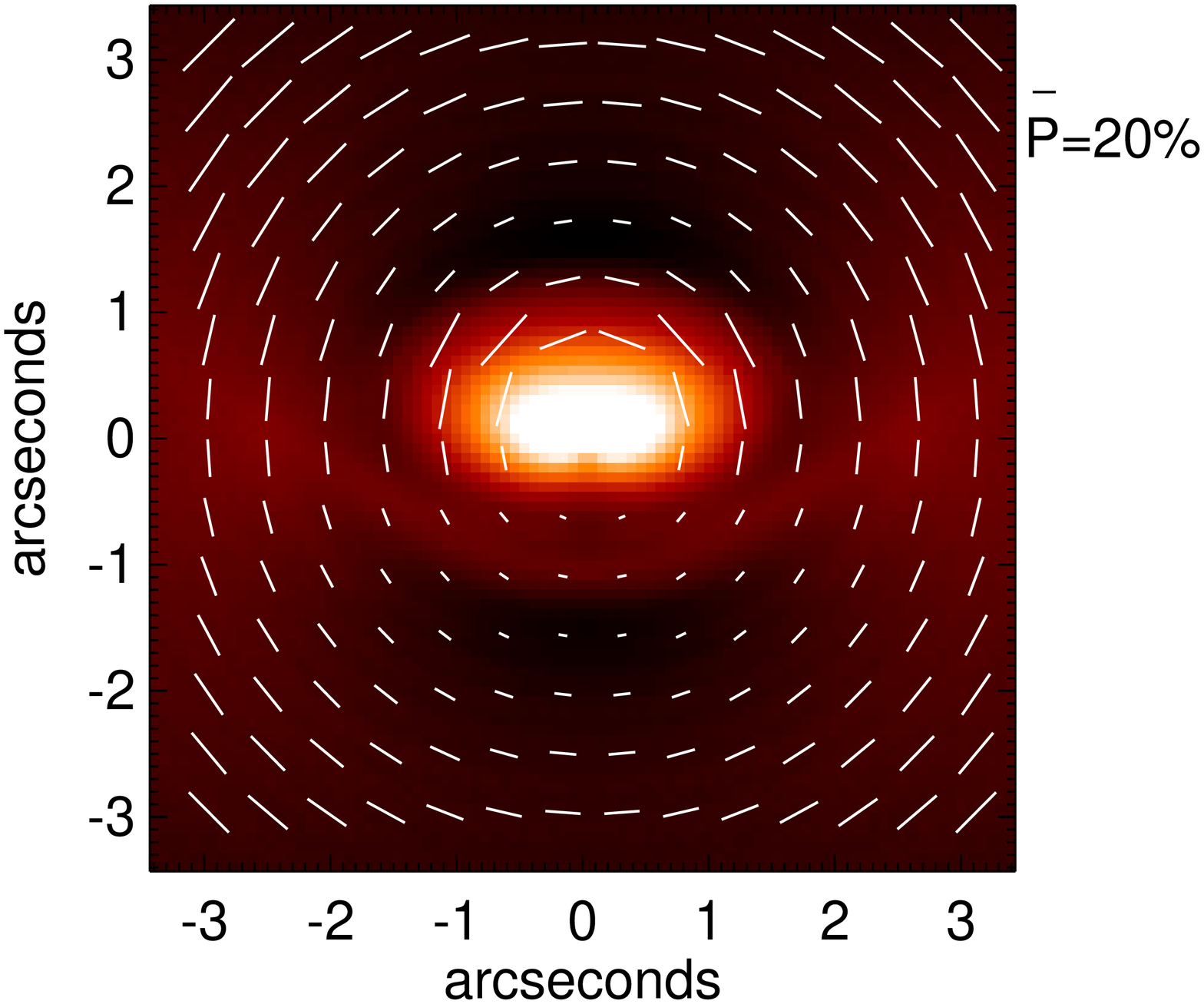}}&
  \imagetop{\includegraphics[width=0.34\textwidth]{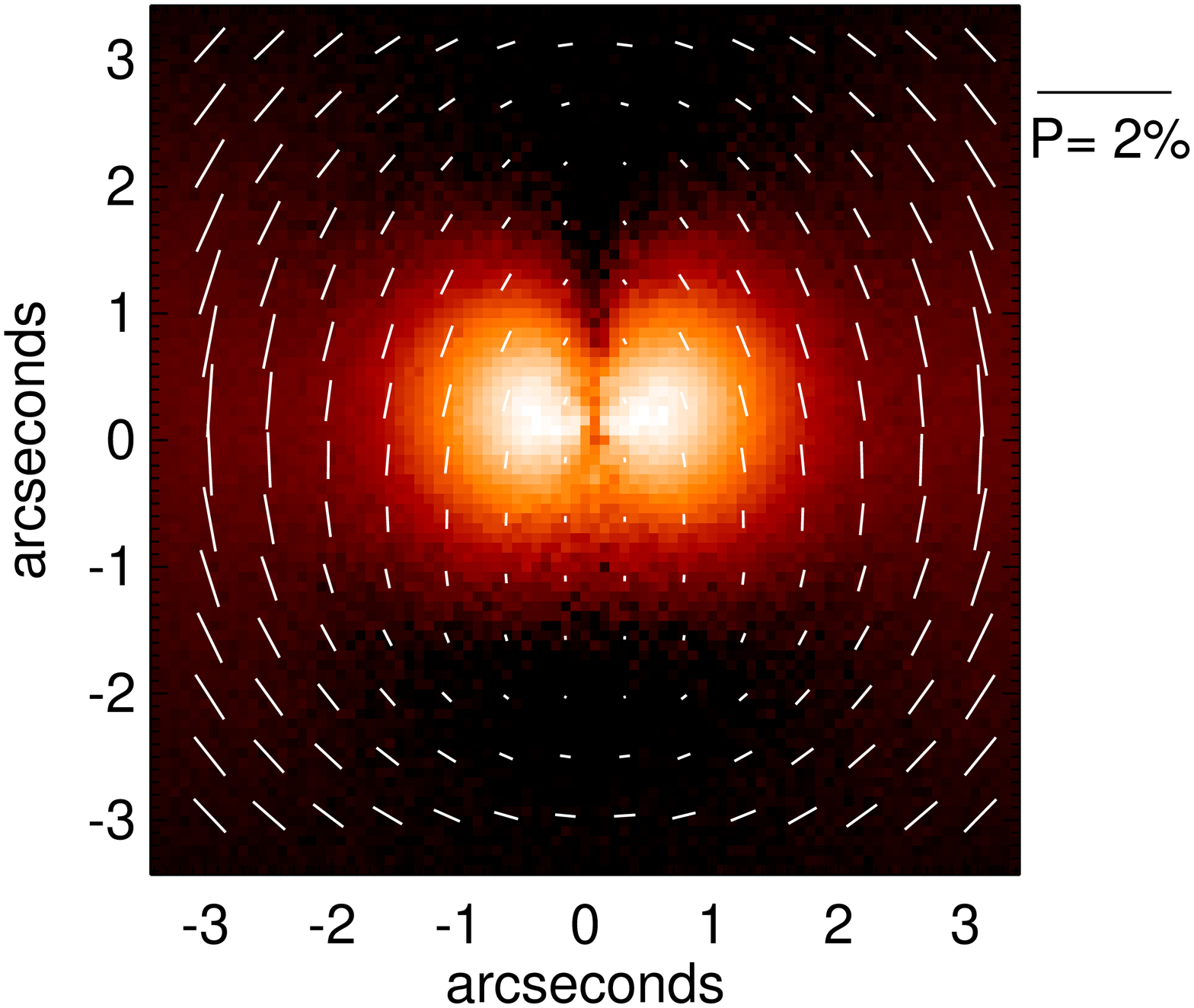}} \\
  \imagetop{\includegraphics[width=0.265\textwidth]{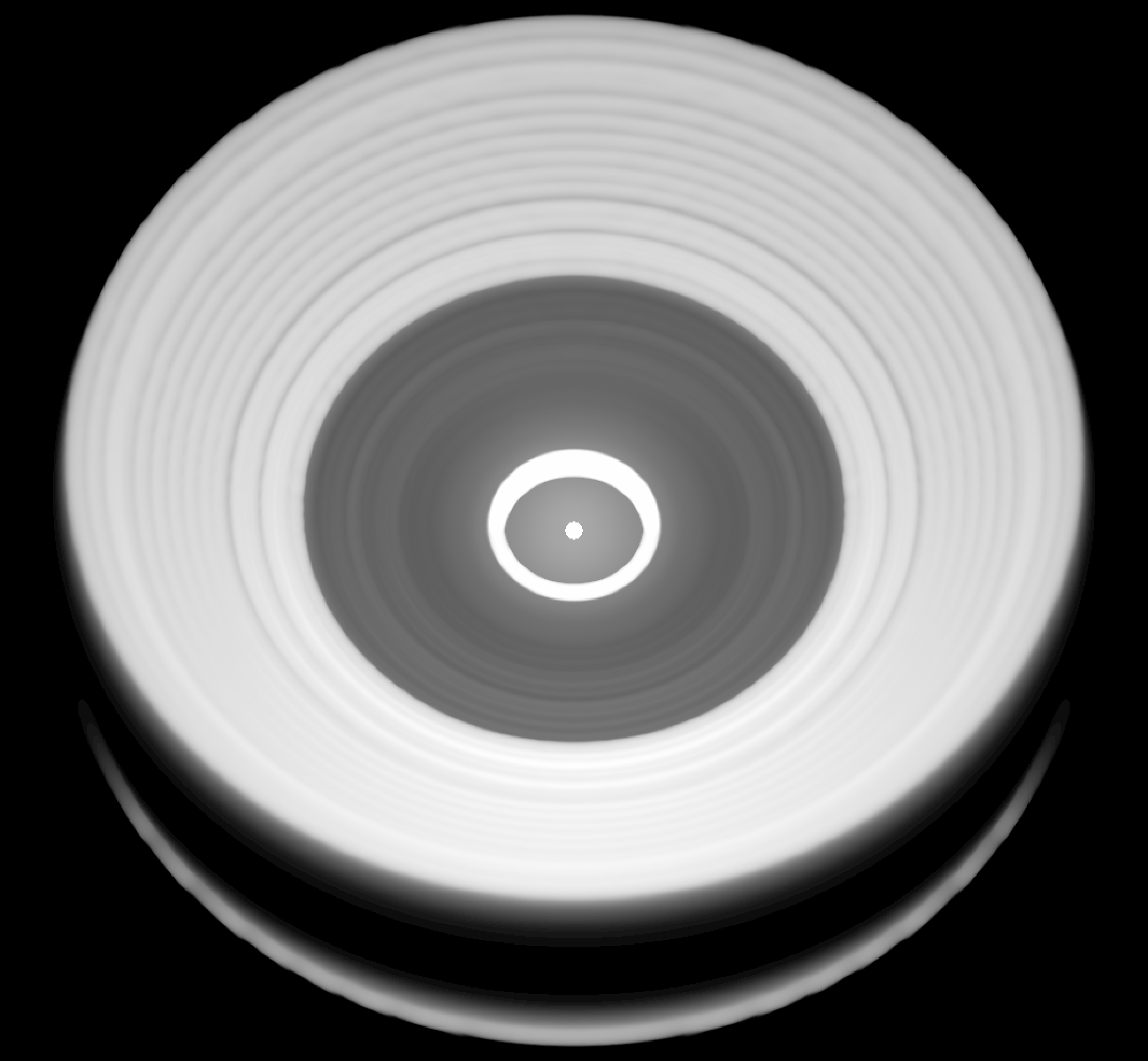}} &
  \imagetop{\includegraphics[width=0.34\textwidth]{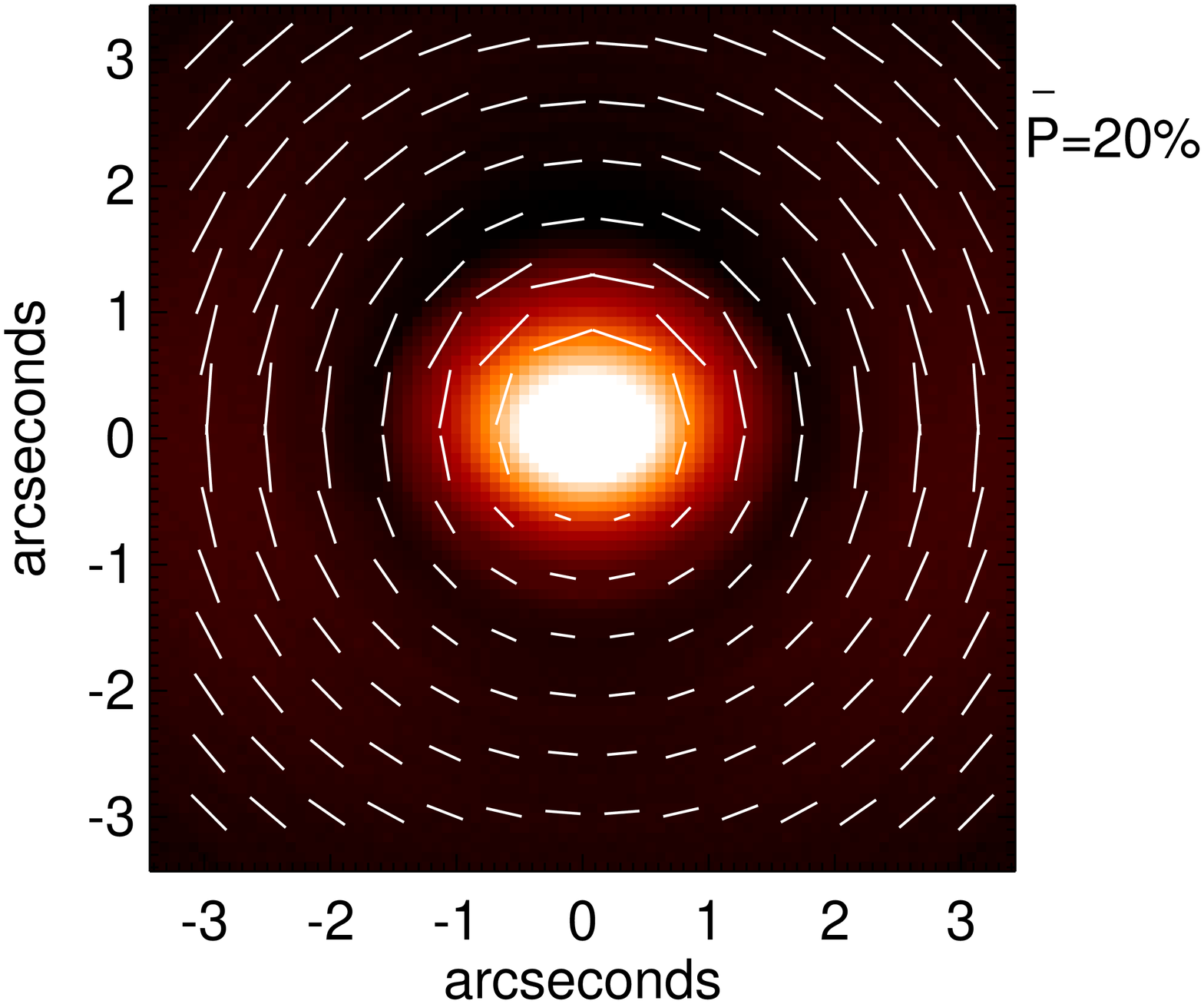}}&
  \imagetop{\includegraphics[width=0.34\textwidth]{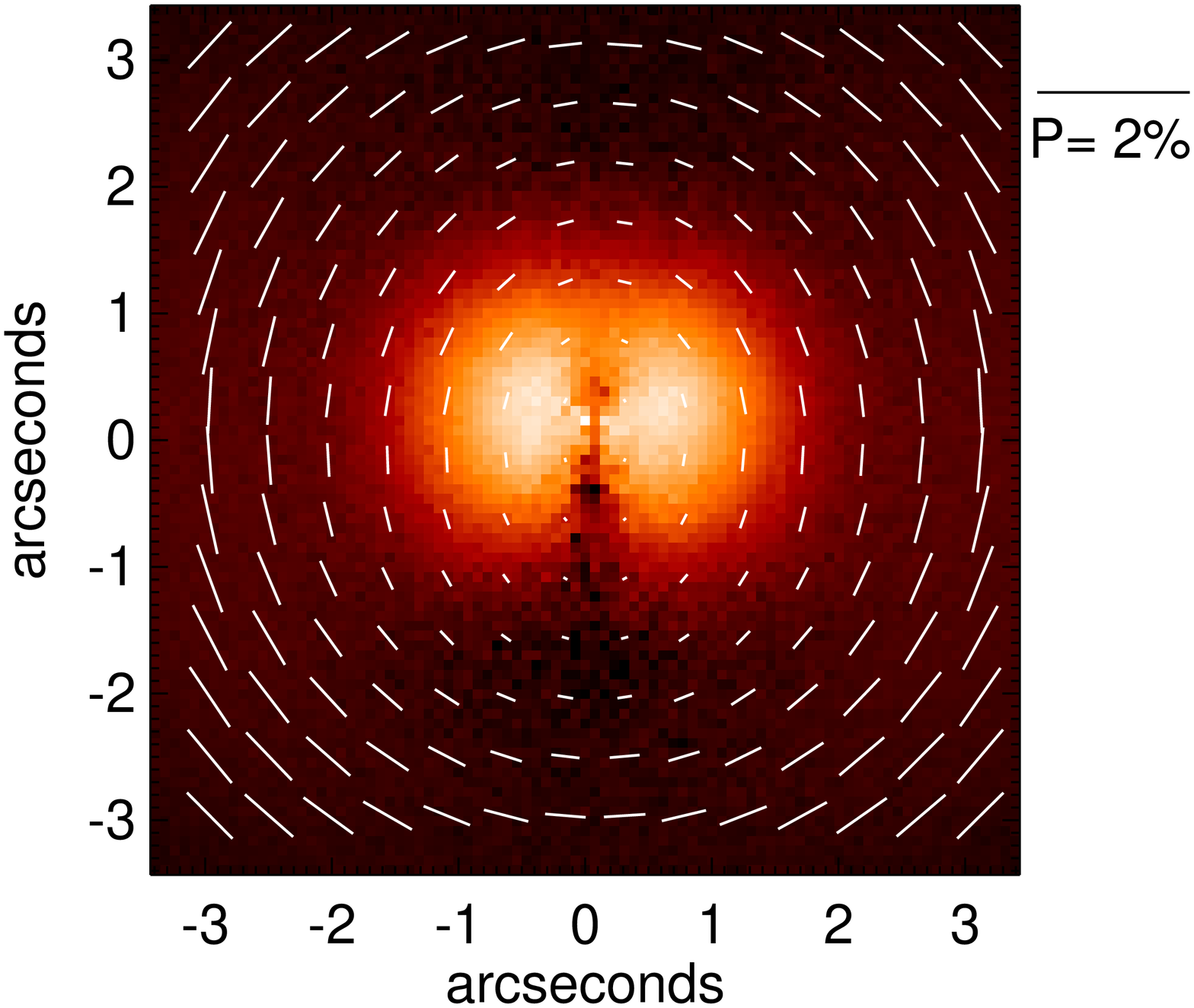}} \\
\end{tabular}
\caption{Models of the disk surrounding SU Aur showing the impact of inclination on the morphology of the disk images in linearly polarised light.  The top panel shows a model inclination of i=90$^\circ$, i=70$^\circ$, i=50$^\circ$, and i=30$^\circ$ in the lowermost panel.  The images in the left-hand panel show the cartoon interpretation of the disk inclination, the images in the middle panel show the raw model images, and the images in the right-hand panel are the polarised intensity models that have been convolved with a simulated ExPo PSF and processed through the ExPo data reduction pipeline.  The intensity scale is logarithmic over two orders of magnitude and is the same as that of the ExPo images.}  
\protect\label{f-expo-inc1} 
\end{figure*}

\begin{figure*} 
\begin{center}
\includegraphics[scale=0.75]{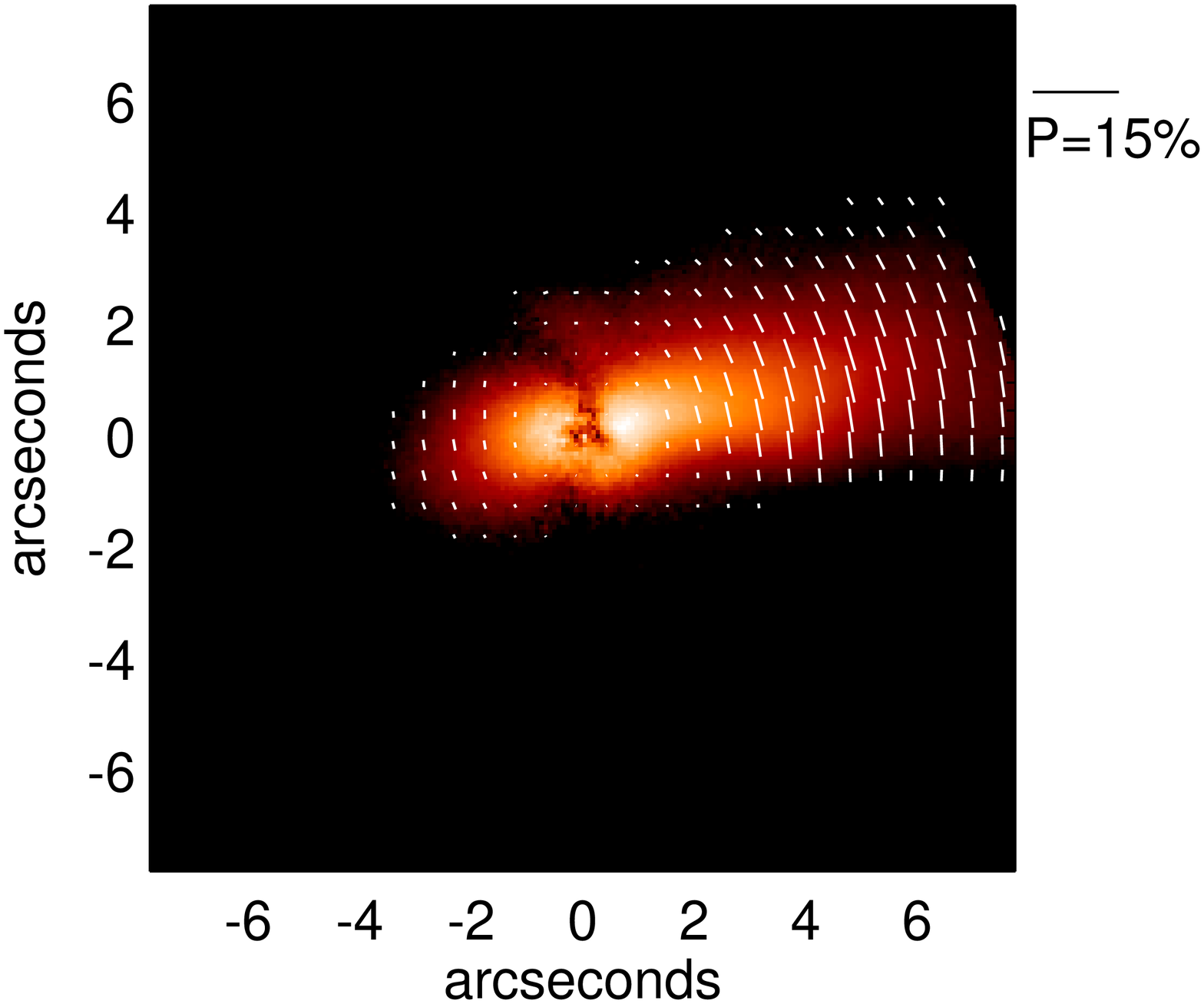}
\caption{Full-field ExPo image of the circumstellar environment of SU Aur in polarised intensity, taken without filter.  The intensity scale is logarithmic over two orders of magnitude and north is up and east is to the left.  The polarisation vectors plotted on the image indicate the fraction of polarised light (P$_I$/I).  At the very centre of the image the vectors are unrealistically small because the total intensity of the star dominates the polarised light. As explained in the text, our derived polarisation degree is a lower limit to the real value.}
\protect\label{f-expo-large} 
\end{center}  
\end{figure*}

To interpret the ExPo images, we constructed a model of the protoplanetary disk surrounding the star using the radiative transfer code MCMax \citep{min2009}. To realistically compare the resulting models with the ExPo images we combined the model images with a simulation of the ExPo instrument that takes into account atmospheric and instrumental effects \citep{Canovas2012,min2012}. We do not claim that the model we present is a unique solution, but we use the model to interpret the large-scale geometrical properties of the disk surrounding SU Aur.

\subsection{Model setup}

The properties of the central star were taken from the literature and were modelled using a Kurucz model with an effective temperature of 5860\,K, a luminosity of 6.3\,L$_{\sun}$, and a mass of $1.88\,M_{\sun}$ at a distance of 143\,pc. We assumed a standard interstellar reddening of the star using $A_V = 0.9$ mag.  The models for SU Aur's disk are based on the models from \citet{min2012}, which were adapted to fit the spectral energy distribution and to be consistent with the ExPo images. The composition of the grains is a mixture of magnesium-rich pyroxene, carbon and iron sulfide in a volume ratio of 75\%, 15\%, and 10\%. The materials were mixed together using the Bruggeman effective medium theory with the addition of 25\% porosity. The refractive-index data were taken from \citet{Dorschner1995}, \citet{Preibisch1993}, and \citet{Begemann1994} for the pyroxene, carbon, and iron sulfide materials, respectively.  To compute the optical properties we used the particle shape model from \citet{min2005} with the irregularity parameter $f_{max}$=0.8.  We used irregularly shaped grains to avoid the unrealistic polarisation behaviour present in the computations of perfect, homogeneous, spherical particles. The size distribution ranges from $0.05\,\mu$m to $a_\mathrm{max}$, via the power law $n(a)da$ $\alpha$ $a^{-p}$, where a$_{max}$ and $p$ are fitting parameters.  For the surface density distribution of the material in the disk, $\Sigma$, we used a modified version of the similarity relation from \citet{hughes2008},
\begin{equation}
\Sigma(R) \propto R^{-\gamma}\exp \left\{- \frac{R}{R_0}\right\},
\label{e-hughes}
\end{equation}
where $R$ is the disk radius and $\gamma$ is the exponent of the power law, which is a fitting parameter at small radii, and $R_0$ is the turnover point at the outer edge of the disk fixed at $500\,$au because lower values of the disk diameter resulted in an model image that was significantly smaller than the ExPo image.  

To reproduce the large near infrared excess, which is most likely caused by accretion luminosity, \citep[e.g.][]{kurosawa2005}, an optically thick gas disk was added inside the dust sublimation radius. This gas disk was treated as a simple accretion disk with a temperature distribution given by \cite{Pringle1981}
\begin{equation}
T=\left\{ \frac{3 G M_\star \dot{M}}{8\pi R^3 \sigma} \left[1-\sqrt{\frac{R_\star}{R}}\right] \right\}^{1/4} 
\end{equation}
This disk extends inwards to two stellar radii, where we assumed that the material is channeled onto the star via magnetospheric accretion \citep{shu1994}, and that the remaining energy is emitted at UV wavelengths. The grain sizes are obtained from fitting the dominant silicate feature at $\sim$10 $\mu$m in SU Aur's Spitzer spectrum  simultaneously with SU Aur's SED (see Figure ~\ref{f-sed}).  The result was then compared with the ExPo images.

Our disk model is in vertical hydrostatic equilibrium, where we solved for the settling of the dust grains self-consistently using the description of \citet{Dubrulle1995}. We fitted the SED of SU Aur using a genetic fitting algorithm. Table~\ref{tab:parameters} lists the fitting parameters and the parameters that were fixed along with the resulting best-fit values. 

\begin{table}[!tbp]
\caption{Values for the model parameters.}
\begin{center}
\begin{tabular}{llc}
\hline
Parameter                    &    Symbol        &    Value \\
\hline
Input parameters \\
\hline
Mass (star)            			&    $M_\star$            	&    $1.88\,M_{\sun}$ \\
Temperature (effective)        	&    $T_\mathrm{eff}$    	&    $5860\,$K \\
Luminosity (star)            		&    $L_\star$            		&    $6.3\,L_{\sun}$ \\
Distance            			&    $D$                		&    $144\,$parsec \\
Surface density turnover point 	&    $R_0$            		&    $500\,$au \\
Minimum grain size            	&    $a_\mathrm{min}$    	&    $0.05\,\mu$m\\
Gas-to-dust ratio            		&    $f_{g/d}$            		&    $120$\\
Inner radius of the gas disk    	&    $R_\mathrm{gas,in}$	&    $2\,R_\star$\\
Mass-infall rate of the cloud$^1$	&    $\dot{M}_\mathrm{infall}$	&    $10^{-7}\,M_{\sun}/$yr\\
Polar opening of the cloud     	&    $\mu_0$            		&    $0.8$\\
\hline
Fitted parameters\\
\hline
Inner radius of the dust disk  	&    $R_\mathrm{dust,in}$	&    $0.17\,$AU\\
Dust mass in the disk        	&    $M_\mathrm{dust}$  	&    $1.2\cdot10^{-4}\,M_{\sun}$ \\
Surface density power-law      	&    $\gamma$        		&    $2.4$\\
Maximum grain size          		&    $a_\mathrm{max}$  	&    $66\,\mu$m\\
Power law of the size distribution	&    $p$              	&    $4.8$\\
Turbulence parameter     		&    $\alpha$            		&    $9\cdot10^{-3}$\\
Mass-accretion rate       		&    $\dot{M}$            	&    $3\cdot10^{-7}\,M_{\sun}/$yr\\
\hline
Inferred parameters \\
\hline
Disk inclination (approximate)	& $i$ & 50 degrees \\
Disk position angle of short axis & & 10-20 degrees \\ 
\hline
\end{tabular}
\end{center}
$^1$ The mass-infall rate is taken from \cite{akeson2005}. 
\label{tab:parameters}
\end{table}

\subsection{Data analysis simulation}

To account for the instrumental and data reduction effects, we combined the model images with a model of the ExPo instrument.  To do this, we generated a set of statistically independent point spread functions (PSFs), with the same exposure time,  average full-width-half maximum (FWHM), and pixel size as the observations discussed here. These PSFs were then convolved with the model images. Photon noise, read-out noise and instrumental polariastion were added afterwards. These simulated images were then reduced with the same pipeline as the real observations. This procedure is described in more detail by \cite{min2012}.  For comparison, raw model images are shown in the middle panel of Figure~\ref{f-expo-inc1}.

\subsection{Extended circumstellar environment}

The large-scale ExPo image (15$\arcsec \times 15 \arcsec$), as shown in Figure~\ref{f-expo-large}, clearly shows a nebulosity extending to the west of the image.  Similar structures around SU Aur have previously been reported by \cite{grady2001},\cite{woodgate2003}, and \cite{nakajima1995}.  Since SU Aur is a classical T Tauri star that is still heavily accreting, the most probable interpretation is that SU Aur is still surrounded by the remnants of its prenatal molecular cloud.   To incorporate the effects of an infalling remnant molecular cloud in our model we added an infalling envelope using the density structure proposed by \citet{whitney1993}, with a mass-accretion rate of $3\cdot10^{-7}M_{\sun}/$yr \citep{akeson2005}.  We let the cloud fall onto the disk at the exponential turnover radius, $R_0$ and opened up a polar cavity with $\mu_0=0.8$ \citep[see][for a description of these parameters]{whitney1993}.  The resulting modelled molecular cloud is very faint, and it was not possible to reproduce the almost constant linear polarised intensity level from disk to nebulosity present in the large-scale ExPo image.  Consequently, the modelled observations derived from the disk models with and without a remnant molecular cloud do not differ significantly.


\begin{figure} 
\includegraphics[scale=0.3]{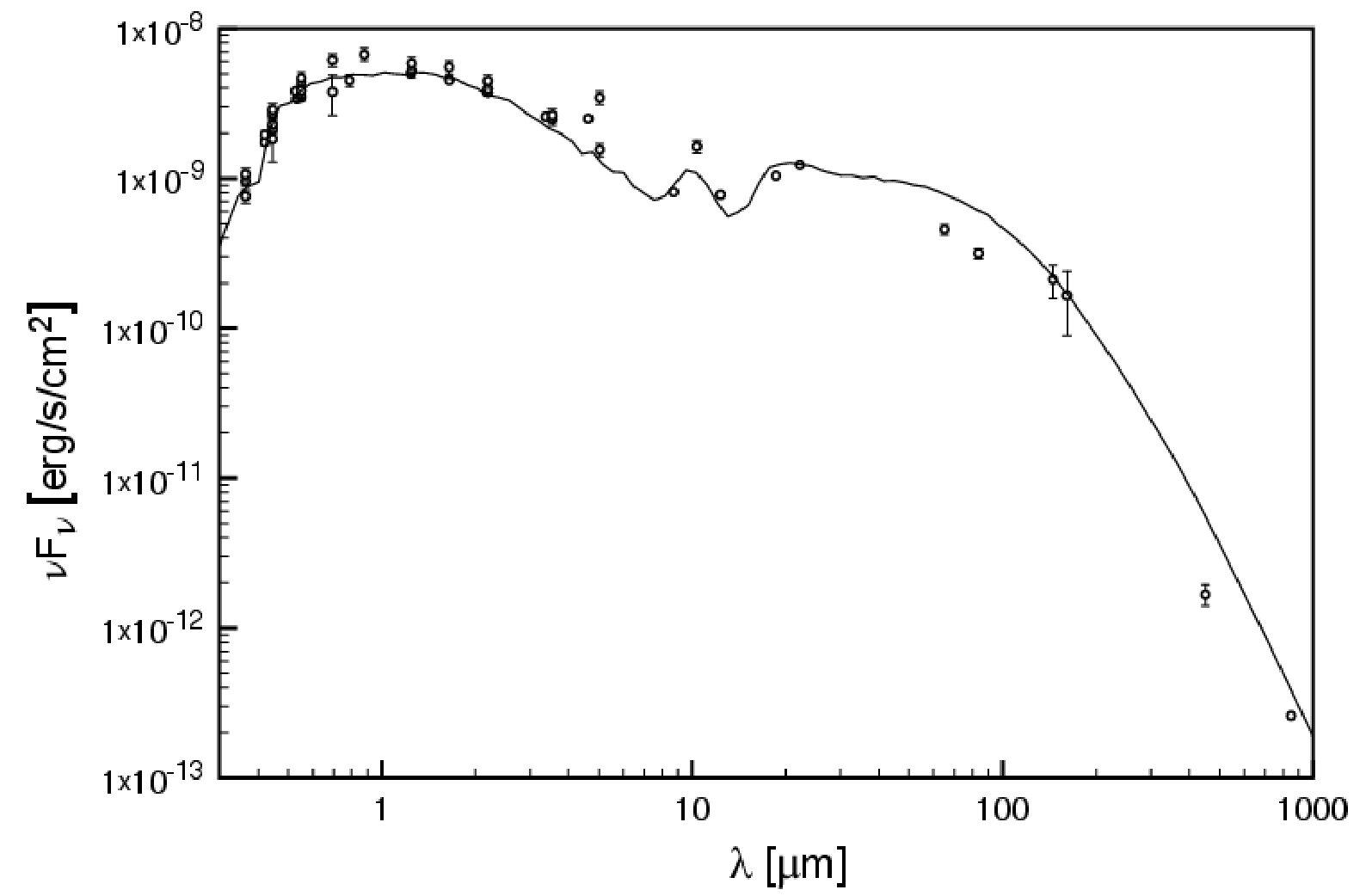}]
\caption{Fitted SED of SU AUR using the model parameters described in Section 5.  Data are taken from \protect\cite{Cutri2012} (WISE), \protect\cite{hog2000} (the Tycho-2 Catalogue), \protect\cite{Mermilliod1991} (Johnson UBV), \protect\cite{Morel1978} (Johnson UBVRIJKLM), \protect\cite{Ishihara2010} (AKARI/IRC mid-IR all-sky Survey - L18W+S9W), \protect\cite{Ducati2002} (Johnson VJKLMN), \protect\cite{Ofek2008} (Johnson JKH), \protect\cite{ESA1997} (Tycho-2.BT+.VT), \protect\cite{Droege2006} (Cousins I, Johnson V), \protect\cite{Monet2003} (USNO-B1.0.B1+.R1), \protect\cite{ISASJAXA2010} (AKARI/IRC mid-IR all-sky Survey - WIDEL+WIDES+N160+N60), \protect\cite{Kharchenko2001} (Johnson BV), \protect\cite{Skrutskie2006} (2MASS JKsH) and \protect\cite{Hauck1990} (Johnson UBV), \protect\cite{Andrews2005} (450 and 850$\mu$m filters).}
\protect\label{f-sed} 
\end{figure}

\subsection{Disk Model}

We determined that the observed SED and ExPo images are best reproduced using a disk model with small dust grains that follow a very steep size-distribution power law.  This implies that the size distribution is dominated by submicron-sized particles with almost no large dust grains and the upper grain size is poorly constrained. The disk extends between 0.16 au and 500 au where the total dust mass is computed to be $1.2\cdot10^{-4}\,M_{\sun}$.  The mass accretion rate of disk material onto the star is $3\cdot10^{-7}\,M_{\sun}/$yr with a 
a turbulence parameter $\alpha$ = $9\cdot10^{-3}$ and a surface-density power law (as a function of radius) $\gamma$ = 2.4 (ref Equation~\ref{e-hughes}).

In Figure~\ref{f-expo-inc1} we show the resulting model at fixed inclination angles. From these models we conclude that the model with an inclination angle of 50$^\circ$ reproduces the observed polarimetric image best (ref Figure~\ref{f-expoim1}). The dark lane through the centre of the polarised intensity image is a result of the subtraction of the polarisation of the central component, which was made to correct any residual instrumental polarisation.  A side effect of this method is that we also subtracted the polarisation originating at the innermost edge of the disk, resulting in the artificial dark lane seen in the image \citep[see also][]{min2012}. Since we used exactly the same data reduction techniques on the observations and the modelled data, the dark lane is also present in the model images that were convolved with the simulated ExPo PSF (Figure~\ref{f-expo-inc1}).  


\section{Discussion}

We have for the first time directly imaged the circumstellar environment of the classical T Tauri star SU Aur.  Using interpretive models, we determined the inner radius of the dust disk, the mass of the dust in the disk, the surface-density power law, the power law of the dust grain size distribution, the turbulence parameter, and the mass-accretion rate.  


\subsection{SU Aur's disk}

The derived disk parameters were computed by fitting the SED and comparing the resulting image with the ExPo images.  This is consequently much more accurate than fitting the SED alone. Additionally, because SU Aur is known to be actively accreting \citep{kurosawa2005}, an optically thick gas disk was added inside the dust-sublimation radius that extended inwards to 2 R$_\star$.  Excluding this additional luminosity source resulted in a very poor fit to the SED at 1.5 to 6 microns.  The resulting fit to the SED corresponds to a simulated image that closely resembles the spatial extent and brightness of the ExPo image. The mass accretion rate of $3\cdot10^{-7}\,M_{\sun}/$yr agrees well with the value computed by \cite{Calvet2004} of $5-6\cdot10^{-7}\,M_{\sun}/$yr.  Additionally, the determined dust disk mass of $1.2\cdot10^{-4}$ M$_\odot$ (assuming a gas-to-dust ratio of 120, ref. Table~\ref{tab:parameters}) is quite typical and is approximately 1 percent of the stellar mass \citep{Scholz2006,williams2011}.

The grain sizes in the surface layers of SU Aur's disk are mostly in the form of small grains that follow a size distribution with a power-law exponent of 4.8.  This implies that most of the mass is in the form of small grains, and because there are very few large grains, their maximum size cannot be clearly constrained.  The grain sizes were obtained from fitting the dominant silicate feature at $\sim$10 $\mu$m in SU Aur's Spitzer spectrum  simultaneously with SU Aur's SED (see Figure ~\ref{f-sed}). Supporting the presence of small grains in the outer layers of SU Aur is the work of \cite{KesslerSilacci2006}, who analysed Spitzer spectra of disks around T Tauri stars.  \cite{KesslerSilacci2006} concluded that the 10$\mu$m silicate feature is uncorrelated with age or the H$_\alpha$ equivalent width, indicating that turbulence and small grain regeneration are important processes in disk evolution.   This is confirmed with recent results from Herschel ~\citep{Oliveira2013} and models of aggregate fragmentation \citep{Dullemond2005,Birnstiel2012}.  More detailed information on the grain sizes, derived from modelling observed images at different wavelengths is not possible given the unstable atmospheric seeing of our observations in different filters.

Previous models of the circumstellar disk surrounding SU Aur agree well with the results of our work.  The models of \cite{akeson2002,akeson2005}, based on interferometric data from infrared and millimeter wavelengths and infrared photometry, showed that SU AUR's disk can be fitted by a passive flat-disk model, which agrees well with the model used in this analysis for the middle to outer regions of SU Aur's disk.  We determined the inclination of the disk to be approximately 50$^\circ$ which is in line with the value of 62$^\circ$ determined by \cite{akeson2005}. The disk position angle of the short axis is inferred from the model images to be between 10$^\circ$ and 20$^\circ$ which matches the position axis of the long axis of 112$\pm$24$^\circ$ determined by \cite{akeson2005}.

The inner radius of the dust disk of 0.17 au agrees well with the value of 0.13/0.18 au modelled by \cite{akeson2002}.  The outer disk radius computed by \cite{akeson2002} (derived to be between 70 and 240 au) using submillimeter and millimeter interferometric observations, is smaller than the outer disk radius of 500au derived in this paper.  An important consideration in comparing the results of the disk size is that the outer edge of our modelled disk is smooth, that is, it is not a sharp cut-off and could result in a wavelength dependence of the outer disk radius.  Moreover, the very high surface density power-law exponent means that there is less mass in the outer regions of the disk.  The outer disk radius of 500 au used in this paper is significantly higher than the value of 100-300 au adopted by \cite{ricci2010} to model the SED of SU Aur.  Without knowing the true spatial extent of the disk, they showed that the submm and mm SED can be fitted with a very compact disk (R$_{out} \approx$ 20 au and $i \approx$ 70$^\circ$), which highlights the importance of including the spatial information derived from the ExPo image in our analysis. 

\begin{figure} 
\includegraphics[scale=0.35,angle=270]{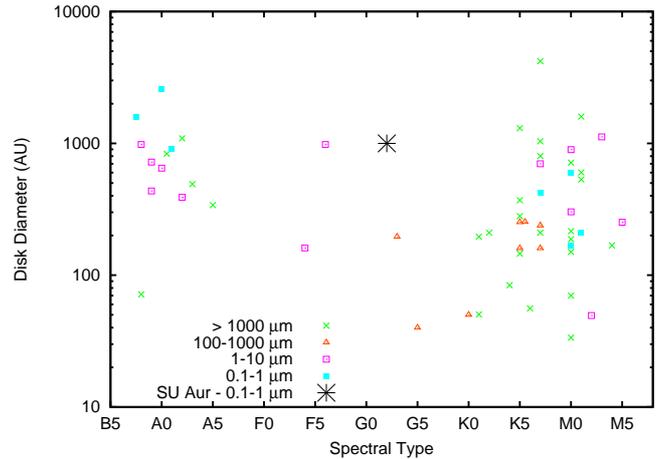}
\caption{Diameter of imaged disks as a function of spectral type, where each observed wavelength range is indicated by a different colour and symbol shape.  SU Aur is shown as the large black star in the centre of the plot.  Data are taken from www.circumstellardisks.org.}  
\protect\label{f-disk-size} 
\end{figure}
   
The turbulence parameter, $\alpha$, was determined from the SED to be $\sim$10$^{-2}$. This value of $\alpha$ is significantly higher than the median value found by \citet{mulders2012} of $\alpha$ = 10$^{-4}$ for both T Tauri and Herbig stars also computed using the MCMAX radiative transfer code to fit an average stellar SED.  The turbulence parameter is important because it defines the degree of grain mixing in the disk, which has implications for the settling of grains in the disk mid-plane and consequently for the formation of planets.  An higher value of $\alpha$ than other young T Tauri disks, might consequently increase the mixing of small grains in the disk, and increase the collision and growth rate in the disk mid-plane.  Because we only observe the surface layers of SU Aur's disk, it is not possible to make a quantitative statement about the possibilities for forming planets, which can only be achieved with more detailed multi-wavelength imaging.  However, it is significant that the turbulence parameter, and consequently the grain mixing in the disk, is significantly higher than typically found in disks around T Tauri and Herbig stars.


In the context of protoplanetary disks, SU Aur is a particularly interesting example because, with a mass of 1.88 M$_\odot$ it lies on the Herbig Ae/Be and T Tauri transition.  An essential difference between these two types is that T Tauri stars have a much longer PMS evolution (10-100 Myr) than the 10 Myr for Herbig Ae/Be stars due to their lower mass.  SU Aur has the mass of a Herbig star, while having the cooler spectrum of a G2 star.  As SU Aur moves towards the main sequence, it will contract and become hotter, and probably arrive on the main-sequence as a late A-type star.  However, SU Aur is currently similar in stellar structure to the Sun, though with increased mass and increased luminosity.  As it evolves towards the main-sequence, it will contract and loose its thin outer convection zone and cross the T Tauri-Herbig transition.  In comparison with other disks that have been resolved for T Tauri stars, SU Aur's disk, with a diameter of 1000 AU, is most similar in size to that of DL Tau (K7 1036 au), IRAS 04158+2805 (M3, 1120 au), CoKu Tau 1 (M0, 896 au), while it is significantly larger than the disks that have been imaged at submm wavelengths around T Tauri stars of similar spectral type, SR21 (G3, 190 au), and UX Tau (G5 40 au).  The size of SU Aur's disk compared with other Herbig and T Tauri stars is shown in Figure ~\ref{f-disk-size}, where it exceeds the median disk size for a Herbig star, and one of the largest T Tauri disks.

The composition of SU Aur's disk has been investigated by \cite{honda2006} who showed that the disk contains polycyclic aromatic hydrocarbon (PAH) emission features similar to those seen in Herbig Ae/Be stars \citep{meeus2001,acke2004,sloan2005}, though the PAH features are much weaker than in other Herbig stars \citep{keller2008} and are not found in other late-type (T Tauri) stars \citep{furlan2006}.  The presence of weak PAH emission is consistent with the senario of our modelled disk being a flaring disk. In the interpretive model of this paper, the flaring angle was solved self-consistently by assuming vertical hydrostatic equilibrium.  


\subsection{Su Aur's extended circumstellar environment}

In our full-field ExPo image we clearly see a large extended nebulosity as shown in Figure~\ref{f-expo-large}.  This has previously been observed by \cite{Herbig1960}, \cite{grady2001},\cite{woodgate2003} and \cite{nakajima1995}.  The nebulosity or extended circumstellar environment has been interpreted to result from light that is scattered off the cavity walls carved out by a large jet \citep{grady2001}. 
However, the image of \cite{grady2001} was subsequently interpreted by \cite{chakraborty2004}, who infer that the position angle of the disk must be perpendicular to the axis of the jet (or nebulosity, as referred to in this paper), with an inclination of 65$^\circ$ and a position angle of 180$^\circ$.  The direct images of the disk presented here indicate an inclination angle of 50$^\circ$ and a position angle of $\approx$10-20$^\circ$, which agrees with results from interferometry \citep{akeson2005} and the spectropolarimetric results of \cite{vink2005}.  This rules out a polar outflow or jet as the cause of the large-scale nebulosity.  We also found that when we included the results of \cite{wood2001}, where such a feature is a remnant cavity wall carved out by a jet or outflow, that the brightness measurements are not sufficient to match the disk or nebulosity brightness ratio of our ExPo images.   Additionally, \cite{Herbig1960} noted that SU Aur's nebulosity extends to the Herbig star AB Aur that is 3' away.  We conclude that the large nebulosity extending from SU Aur is most likely a remnant of the prenatal cloud that formed both SU Aur and AB Aur.

\section{Conclusions}

We have presented the first direct images of the circumstellar environment of the young classical T Tauri SU Aur at the Herbig Ae/Be and T Tauri transition.  SU Aur's disk was modelled using radiative-transfer models to fit the SED, the results of which were compared with the ExPo image.  We found very small grains in the surface layers of the disk.  The modelled turbulence parameter is about two orders of magnitude higher than previously determined for Herbig Ae/Be and T Tauri disks, indicating an enhanced level of grain mixing in the disk, though this value needs to be confirmed with more detailed multi-wavelength (imaging) studies.  In our ExPo images, we additionally resolved a large extended nebulosity that is most likely a remnant of the prenatal molecular cloud because of the position angle of the disk and the brightness of the nebulosity, and is not a cavity carved out by a jet, as previously speculated.

\section*{Acknowledgments}
We acknowledge support from NWO and thank Pieter Degroote (Leuven), who supplied the observed SED.  S.V.J. acknowledges research funding by the Deutsche Forschungsgemeinschaft (DFG) under grant SFB 963/1, project A16, H.C.C. acknowledges support from the Millenium Science Initiative of the Chilean Ministry of Economy, Nucleus P10-022-F, and M.M. acknowledges funding from the EU FP7-2011 under Grant Agreement nr. 284405. This publication makes use of data products from the Wide-field Infrared Survey Explorer, which is a joint project of the University of California, Los Angeles, and the Jet Propulsion Laboratory/California Institute of Technology, funded by the National Aeronautics and Space Administration. This publication makes use of data products from the Two Micron All Sky Survey, which is a joint project of the University of Massachusetts and the Infrared Processing and Analysis Center/California Institute of Technology, funded by the National Aeronautics and Space Administration and the National Science Foundation.  

\bibliographystyle{aa}
\bibliography{iau_journals,suaur}

\label{lastpage}

\end{document}